\documentclass[letter]{aa} 
\usepackage{graphicx}
\usepackage{txfonts}

\newcommand{\zav}[1]{\left(#1\right)}
\newcommand{\hzav}[1]{\left[#1\right]}

\begin{document}
   \title{
Rotationally modulated variations and the mean longitudinal magnetic field 
of the Herbig Ae star HD 101412\thanks
{Based on observations obtained at the European Southern Observatory (ESO programme 085.C-0137(A)).}}
\titlerunning{The magnetic field of HD\,101412}

   \author{
S.~Hubrig\inst{1}
\and Z.~Mikul\'{a}\v{s}ek\inst{2,3}
\and J.~F.~Gonz\'alez\inst{4}
\and M.~Sch\"oller\inst{5}
\and I.~Ilyin\inst{1}
\and M.~Cur\'e\inst{6}
\and M.~Zejda\inst{2}
\and C.~R.~Cowley\inst{7}
\and V.~G. Elkin\inst{8}
\and M.~A.~Pogodin\inst{9,10}
\and R.~V.~Yudin\inst{9,10}
}
\authorrunning{Hubrig et al.}

\institute{Astrophysikalisches Institut Potsdam, An der Sternwarte 16, 14482 Potsdam, Germany\\
              \email{shubrig@aip.de}
\and Department of Theoretical Physics and Astrophysics, Masaryk University, Brno, Czech Republic
\and Observatory and Planetarium of J. Palisa, V{\v S}B - Technical University, Ostrava, Czech Republic
\and Instituto de Ciencias Astronomicas, de la Tierra, y del Espacio (ICATE), 5400 San Juan, Argentina  
\and European Southern Observatory, Karl-Schwarzschild-Str.\ 2, 85748 Garching bei M\"unchen, Germany        
\and Departamento de F\'isica y Astronom\'ia, Facultad de Ciencias, Universidad de Valpara\'iso, Chile
\and Department of Astronomy, University of Michigan, Ann Arbor, MI 48109-1042, USA
\and Jeremiah Horrocks Institute of Astrophysics, University of Central Lancashire, Preston PR1 2HE, United Kingdom
\and Pulkovo Observatory, Saint-Petersburg, 196140, Russia
\and Isaac Newton Institute of Chile, Saint-Petersburg Branch, Russia
}
\date{Received; accepted}

 
  \abstract
   {
Despite of the importance of magnetic fields for the full understanding of the properties of accreting Herbig Ae/Be stars, 
these fields have scarcely been studied over the rotation cycle until now. One reason for the paucity of such observations
is the lack of knowledge of their rotation periods. 
The sharp-lined young Herbig Ae star HD\,101412 with a strong surface magnetic field became in the 
last years one of the most studied targets among the Herbig Ae/Be stars.
   }
   {
A few months ago we obtained multi-epoch polarimetric spectra of this star with FORS\,2 
to search for a rotation period and to constrain the geometry of the magnetic field.
}
   {We measured longitudinal 
magnetic fields on 13 different epochs distributed over 62 days.
These new measurements together with our previous measurements of the magnetic field in this star 
were combined with available photometric observations
to determine the rotation period. 
}
   {
The search of the rotation period resulted in  $P=42.076 \pm 0.017$\,d. 
According to near-infrared imaging studies the star is observed nearly edge-on.
The star exhibits a  single-wave variation of 
the longitudinal magnetic field during the stellar rotation cycle. These observations are usually considered as evidence 
for a dominant dipolar contribution to the magnetic field topology.
}
   {}

   \keywords{
stars: pre-main-sequence --
stars: atmospheres --
stars: individual: HD\,101412 --
stars: magnetic fields --
stars: rotation --
stars: variables: general }

   \maketitle
%

\section{Introduction}

A number of Herbig Ae stars and classical T\,Tauri stars are surrounded 
by active accretion disks and, probably, most of the excess emission seen at
various wavelength regions can be attributed to the interaction of the disk
with a magnetically active star (e.g.\ Muzerolle et al.\ \cite{Muzerolle2004}).
This interaction is generally referred to as
magnetospheric accretion. Recent magnetospheric accretion models for these stars 
assume a dipolar magnetic field
geometry and accreting gas from a circumstellar disk falling ballistically 
along the field lines onto the stellar surface. 

Despite of the importance of magnetic fields for the full understanding of the properties of accreting Herbig Ae/Be stars, 
these fields have scarcely been studied over the rotation cycle until now. One reason for the paucity of such observations
is the lack of knowledge of their rotation periods. The rotational period has not been known for any Herbig Ae star or
debris disk star.
The $v_{\rm eq}$ values can be estimated for stars with known disk inclinations and $v \sin i$,
and from the knowledge of $v_{\rm eq}$ values and stellar radii the rotation periods can be deduced.
Our estimation of rotation periods of 17 Herbig Ae stars and four debris disk stars showed that 
for a major part of the studied sample the periods are of the order of a few days or fractions of days, 
and only for one Herbig Ae star, HD\,101412, the expected period was 
longer than 17 days (Hubrig et al.\ \cite{Hubrig09}). 
Among the studied sample, HD\,101412 also showed the largest longitudinal magnetic 
field, $\left<B_{\rm z}\right>$\,=\,$-$454$\pm$42\,G, measured on low-resolution polarimetric spectra
obtained with FORS\,1 (FOcal Reducer 
low-dispersion Spectrograph) mounted on the 8-m Kueyen (UT2) telescope of the VLT.
The subsequent spectroscopic study of twelve UVES and HARPS spectra of HD\,101412 revealed the presence of 
resolved magnetically 
split lines indicating a variable magnetic field modulus changing  from 2.5 to 3.5\,kG
(Hubrig et al.\ \cite{Hubrig10}). 
The presence of a rather strong magnetic field on the surface of HD\,101412 makes it a prime 
candidate for studies of the impact of the magnetic field on the physical processes occurring during stellar formation.
In this 
work we present for the first time a mean longitudinal magnetic field measurement 
series, obtained with the multi-mode instrument FORS\,2  at the VLT.
The data and available photometric observations are used to obtain the rotation period and to put 
constraints on the magnetic field geometry. This is the first, necessary step for future
more detailed magnetic studies of this remarkable star.

\section{Period determination}


Multi-epoch series of polarimetric spectra of the Herbig Ae star HD\,101412 were obtained with FORS\,2\footnote{
The spectropolarimetric capabilities of FORS\,1 were moved to
FORS\,2 in 2009.
}
on Antu (UT1) from 2010 March 30 to 2010 June 1 in service mode.
Using a slit width of 0\farcs4 the achieved spectral resolving power 
of FORS\,2 obtained with the GRISM 600B was about 2000.
 A detailed description of the assessment of the longitudinal 
magnetic field measurements using FORS\,2 is presented in our previous papers 
(e.g., Hubrig et al.\ \cite{Hubrig2004a, Hubrig2004b}, and references therein). 
The mean longitudinal 
magnetic field, $\left< B_{\rm z}\right>$, was derived using 

\begin{equation} 
\frac{V}{I} = -\frac{g_{\rm eff} e \lambda^2}{4\pi{}m_ec^2}\ \frac{1}{I}\ 
\frac{{\rm d}I}{{\rm d}\lambda} \left<B_{\rm z}\right>, 
\label{eqn:one} 
\end{equation} 

\noindent 
where $V$ is the Stokes parameter which measures the circular polarisation, $I$ 
is the intensity in the unpolarised spectrum, $g_{\rm eff}$ is the effective 
Land\'e factor, $e$ is the electron charge, $\lambda$ is the wavelength, $m_e$ the 
electron mass, $c$ the speed of light, ${{\rm d}I/{\rm d}\lambda}$ is the 
derivative of Stokes $I$, and $\left<B_{\rm z}\right>$ is the mean longitudinal magnetic 
field. 
 
 

An initial frequency analysis  was performed  on the longitudinal magnetic field 
measurements using a non-linear least-squares fit of the multiple harmonics utilizing the Levenberg-Marquardt 
method (Press et al.\ \cite{Press92}) with an optional possibility of pre-whitening of the trial harmonics.  To detect 
the most probable period we calculated the frequency spectrum for the same harmonic with a number of trial 
frequencies by solving the linear least-squares problem. At each trial frequency we performed a statistical 
test of the null hypothesis for the absence of periodicity (Seber \cite{Seber77}), i.e.\ testing that all harmonic 
amplitudes are at zero.
The resulting amplitude spectrum clearly showed a dominant peak with an equivalent period of 42.098\,d.

To increase the accuracy of the rotation period determination and to exclude the presence of other periods,
we used photometric data that include 920 $V$-band 
and 539 $I$-band observations from ASAS (Pojma\'nski \cite{pojm02}) and 1426 $R$-band observations from the 
robotic telescope Pi of the Sky (Burd et al. \cite{burd05}; Malek et al. \cite{malek10}).
Further, 78 triads of measurements ($\mathit{UBV}$ bands) over eight nights with internal
accuracy of 9.4, 6.4, and 6.6\,mmag were obtained in
South Africa Astrophysical Observatory (SAAO) in April 2010. 
Combining the magnetic and photometric data, and using the method of Mikul\'{a}\v{s}ek et al. (\cite{zdenek10}), we obtain 

\begin{equation}
\left<V\&I\right>^{\rm max} = {\rm MJD}52797.4 \pm 0.8 + 42.076 \pm 0.017 E
\end{equation}

\begin{figure}
\centering
\includegraphics[width=0.40\textwidth]{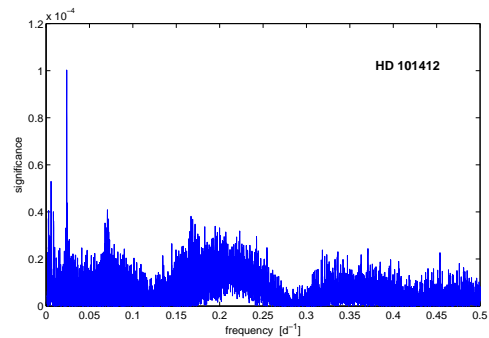}
\caption{
Periodogram of HD\,101412 built from $VI$ ASAS data, data
from Pi of the Sky, SAAO $UBV$, and $\left<B_z\right>$ data.
The periodogram displays only one prominent frequency $f=0.0237665$\,d$^{-1}$, corresponding to
the stellar rotational period of 42.076 days.} 
\label{fig:period}
\end{figure}

\begin{figure}
\centering
\includegraphics[width=0.40\textwidth]{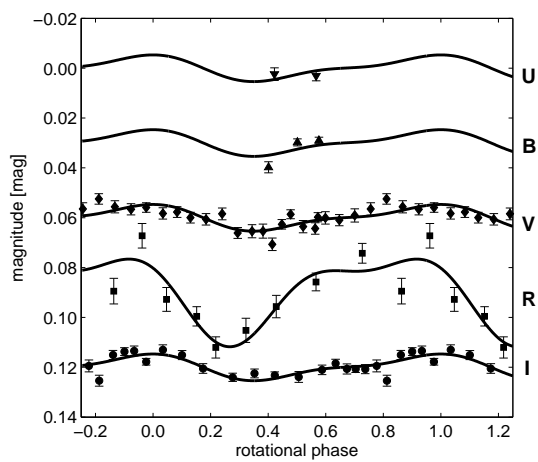}
\caption{
$U\,B\,V\,R$, and $I$ light curves of HD\,101412. Solid lines present the simplest periodic functions
representing the observed photometric variations (see Appendix A). Because of the
large scatter of individual measurements (see Appendix B) the standard technique of normal points (as averages of 
several tens of phase adjacent
measurements) was used. The light curves are shifted along the y-axis for clarity. 
}
\label{fig:krivky}
\end{figure}

The corresponding periodogram and the light curve variations in the
$U$, $B$, $V$, $R$, and $I$ bands are presented 
in Figs.~\ref{fig:period} and ~\ref{fig:krivky}.
The full description of the photometric analysis is given in Appendices A and B.
\begin{figure}
\centering
\includegraphics[angle=270,width=0.40\textwidth]{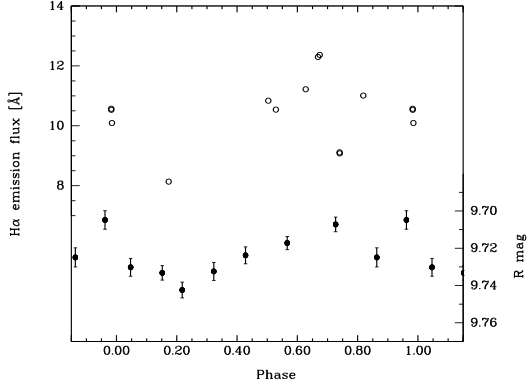}
\caption{
Variations of the H$\alpha$ emission flux over the rotation period (open circles), compared to the variation of the 
$R$-band light curve (filled circles). The internal errors of the Halpha emission flux measurements are 
ordinarily smaller that the symbol size, but gross variations of the flux from
cycle to cycle are expected.}
\label{fig:phaseEW}
\end{figure}
A very asymmetrical shape of the $R$-band light curve, with a deeper minimum at phases 0.15--0.30 and 
large scatter at the phase 0.85 is probably related to a specific wavelength range covered 
by this band, which includes the H$\alpha$ emission line.
As we showed in our previous work, this line is doubled-peaked and strongly variable (see e.g.\ Fig.~1 in 
Hubrig et al.\ \cite{Hubrig10}). The radial velocity of the blue component changes from $-$33.6 to
$-$67.5\,km\,s$^{-1}$ whereas the red component, which actually appears triple in the phase 0.17, is shifted by 
25.5 to 119.6\,km\,s$^{-1}$.
Our measurements of the H$\alpha$ emission line flux over the rotation cycle on available HARPS 
and UVES spectra presented in Fig.~\ref{fig:phaseEW} do not contradict this proposition. However,  
the secondary minimum in the emission flux appears at phase 0.74. The phases of the minima in the measured values 
roughly coincide with the phases where the longitudinal field approaches zero value, i.e.\ close to the 
magnetic equator.
\begin{table}
\centering
\caption{
Magnetic field measurements of HD\,101412 with FORS\,2.
Phases are calculated according to the ephemeris of 
${\rm MJD} = 52797.4 + 42.076~{\rm E}$.
All quoted errors are 1$\sigma$ uncertainties.
}
\label{tab:log_meas}
\begin{tabular}{ccr@{$\pm$}lr@{$\pm$}l}
\hline
\multicolumn{1}{c}{MJD} &
\multicolumn{1}{c}{Phase} &
\multicolumn{2}{c}{$\left<B_{\rm z}\right>_{\rm all}$} &
\multicolumn{2}{c}{$\left<B_{\rm z}\right>_{\rm hyd}$} \\
 &
 &
\multicolumn{2}{c}{[G]} &
\multicolumn{2}{c}{[G]} \\
\hline
\hline
54609.190 & 0.060 & $-$312 & 32 & $-$454 & 42 \\
54610.081 & 0.081 & $-$235 & 28 & $-$317 & 35 \\
55285.141 & 0.125 & $-$132 & 54 & $-$149 & 68 \\
55286.115 & 0.148 &     25 & 48 &    209 & 74 \\
55287.107 & 0.172 &     49 & 36 &     96 & 61 \\
55311.142 & 0.743 & $-$202 & 44 & $-$235 & 56 \\
55312.068 & 0.765 & $-$206 & 48 & $-$241 & 60 \\
55320.056 & 0.955 & $-$526 & 46 & $-$784 & 57 \\
55322.149 & 0.004 & $-$360 & 50 & $-$568 & 74 \\
55323.224 & 0.030 & $-$393 & 51 & $-$515 & 75 \\
55324.073 & 0.050 & $-$359 & 63 & $-$453 & 64 \\
55327.159 & 0.124 &  $-$25 & 49 &  $-$44 & 59 \\
55334.071 & 0.288 &    238 & 57 &    338 & 66 \\
55335.152 & 0.314 &    351 & 75 &    389 & 84 \\
55348.097 & 0.621 &    168 & 44 &    322 & 60 \\
\hline
\end{tabular}
\end{table}

\begin{figure}
\centering
\includegraphics[width=0.45\textwidth]{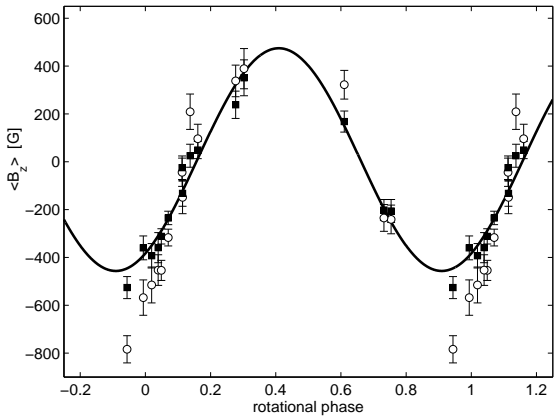}
\caption{
Phase diagram with the best sinusoidal fit for the longitudinal magnetic field measurements using 
all lines (filled squares) and hydrogen lines (open circles). 
}
\label{fig:Bz}
\end{figure}

The logbook of the FORS\,2 spectropolarimetric observations is presented in Table~\ref{tab:log_meas}.
In the first column we show the MJD values for the middle of each spectropolarimetric observation.
The phases of measurements of the magnetic field 
are listed in Column~2. In Columns~3 and 4 we present the longitudinal magnetic 
field $\left<B_{\rm z}\right>_{\rm all}$ using the whole spectrum and the longitudinal magnetic field 
$\left<B_{\rm z}\right>_{\rm hyd}$ using only the hydrogen lines. 
In the first two lines of the table we list the earlier measurements 
published by Hubrig et al.\ (\cite{Hubrig09}).
The corresponding phase diagram for all available mean longitudinal magnetic field measurements
using the whole spectrum and for those using only the hydrogen lines, including the best sinusoidal fit, is
shown in Fig.~\ref{fig:Bz}. The variation has a mean of $\overline{\left< B_{\rm z}\right>} = 9\pm 18$\,G
and an amplitude of $A_{\left< B_{\rm z}\right>} = 465 \pm 27$\,G.
 
\begin{figure}
\centering
\includegraphics[width=0.65\textwidth]{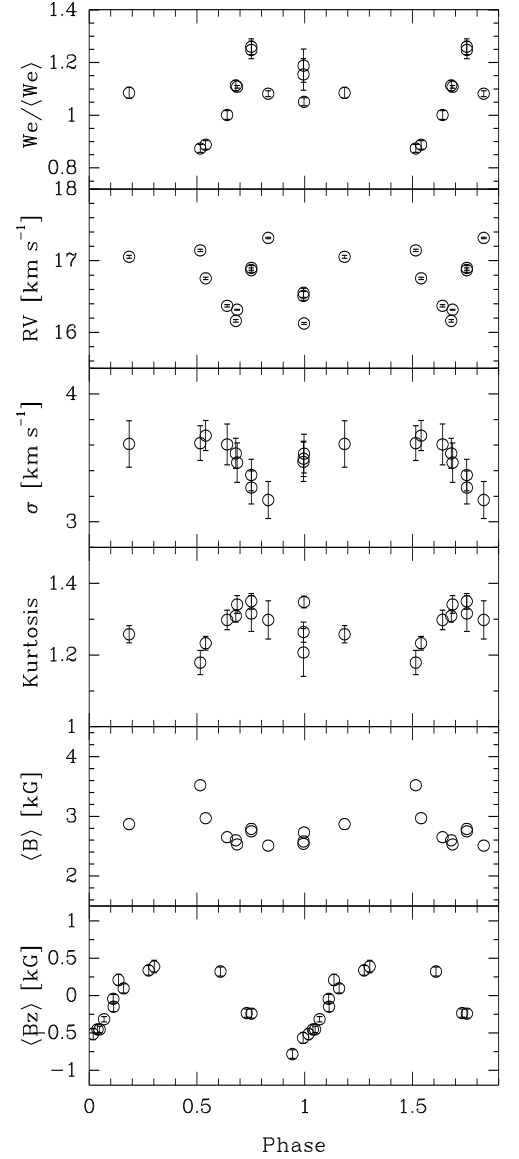}
\caption{
Variations of equivalent width, radial velocity, line width, line asymmetry,
mean magnetic field modulus, and mean longitudinal magnetic field as a function of the rotational phase.
}
\label{fig:6curves}
\end{figure}

In Fig.~\ref{fig:6curves} we present 
the variations of equivalent width, 
radial velocity, line width, and line asymmetry of a sample of iron lines, the
mean magnetic field modulus, and the mean longitudinal magnetic field as a function of the rotational phase.
Due to imperfect phase coverage the interpretation of the observed variations of the iron line profiles is not 
straightforward, but it seems that equivalent width and line asymmetry minima
appear in the phase close to 
the surface field maximum of 3.5\,kG and the positive extremum of the longitudinal field, while the presence
of a secondary minimum is likely at the phase of the negative extremum of the longitudinal field.
The line width is the largest at the phase close to the field positive extremum. The radial velocity variations
are most difficult to interpret: it is possible that they show double modulation, but it is not well-resolved in 
our measurements due to insufficient phase coverage. 

\section{Discussion}

HD\,101412 exhibits a single-wave variation of 
the longitudinal magnetic field during the stellar rotation cycle. These observations are usually considered as evidence 
for a dominant dipolar contribution to the magnetic field topology. 
Our recent study of the abundances of HD\,101412 using UVES and HARPS spectra
 resulted in $v \sin i  = 3\pm 1$\,km\,s$^{-1}$ (Cowley et al.\ \cite{Cowley2010}).
The inclination angles of disks of Herbig Ae/Be stars (which are expected to be identical with the inclination angle
of the stellar rotation axis) can be reliably derived only for resolved observations of disks.
Fedele et al.\ (\cite{Fedele2008}) used VLTI/MIDI observations to determine $i$=80$\pm$7$^{\circ}$.
Using this value we obtain $v_{\rm eq}$=3$\pm$1\,km\,s$^{-1}$. 
Assuming that HD\,101412 is an oblique dipole 
rotator,  we follow the definition of Preston (\cite{Preston1967}): 
 
\begin{equation} 
r = \frac{\left< B_{\rm z}\right>^{\rm min}}{\left< B_{\rm z}\right>^{\rm max}} 
  = \frac{\cos \beta \cos i - \sin \beta \sin i}{\cos \beta \cos i + \sin \beta 
\sin i}, 
\end{equation} 
 
\noindent 
so that the obliquity angle $\beta$ is given by
 
\begin{equation} 
\beta =  \arctan \left[ \left( \frac{1-r}{1+r} \right) \cot i \right]. 
\label{eqn:4} 
\end{equation} 
From the phase curve determined above, with $\left< B_{\rm z} \right>^{\rm max}$=474$\pm$32\,G
and $\left< B_{\rm z} \right>^{\rm min}$=$-$456$\pm$32\,G, we find $r$=$-$0.962$\pm$0.075, 
which for $i$=80$\pm$7$^{\circ}$ leads to a magnetic obliquity of $\beta$=84$\pm$13$^{\circ}$. 

We note, however, that there are some gaps in the phase coverage for the mean longitudinal magnetic field observations,
especially the extrema of the variation curve are not 
very well constrained.  The true magnetic field geometry can probably be resolved once additional longitudinal 
magnetic field measurements as well as magnetic field modulus measurements over the full stellar rotation cycle 
become available in future observations.

For the first time rotation modulated longitudinal magnetic field measurements and
photometric observations were used to determine the rotation period in a Herbig Ae star.
The detection of a large-scale organised predominately dipolar magnetic field on the surface of the young
Herbig Ae star HD\,101412 confirms the scenario that the accretion phenomenon in young stellar objects is 
magnetospheric accretion. However, 
we are still at the very beginning of learning how magnetic fields of Herbig Ae stars
can be incorporated in the modeling
of stellar magnetospheres which takes into account the full complexity of the circumstellar environment, including 
the observed outflows and jets in these objects. 

Longitudinal magnetic fields of the order of a few hundred Gauss 
have  been detected in about a dozen Herbig Ae stars (e.g., Hubrig et al.\ \cite{Hubrig04}, 
\cite{Hubrig06}, \cite{Hubrig2007}, \cite{Hubrig09};
Wade et al.\ \cite{Wade2007}; Catala et al.\ \cite{Catala2007}).
For the majority of these stars rather small fields were measured, of the order of only 
100\,G or less. 
Due to the presence of the rather strong magnetic field 
in the atmosphere of HD\,101412, and a very low  $v \sin i$ value of 3$\pm 1$\,km\,s$^{-1}$,
this star is one of the 
most suitable targets to study its environment, which includes the magnetosphere,
the accretion disk, and the disk wind, altogether producing prominent emission features in the hydrogen line 
profiles.

\begin{acknowledgements}
We acknowledge the support by grants GAAV IAA301630901 and GA{\v C}R  205/08/0003.
\end{acknowledgements}

\section{Appendix A: The search for periodic variations}
The continuously updated archive of the All Sky Automated Survey
(ASAS) is a useful source of photometric observations. 
Observations were obtained from two ASAS
observing stations, one in LCO, Chile (since 1997) and the other on
Haleakala, Maui (since 2006). Both are equipped with two wide-field
instruments, observing simultaneously in $V$ and $I$ band.
More details and the data archive are on
http://www.astrouw.edu.pl/asas/.
After removing observations of declared low quality (C, D) and
apparent outliers, we obtained a set of 539 $I$-band and 920 $V$-band measurements
well covering the periods 1997--1999 and 2000--2009, respectively.
The robotic telescope Pi of the Sky  has been
designed to monitor a significant fraction of the sky with
good time resolution. The final detector consists of two sets
of 16 cameras, one camera covering a field of view of 20$^\circ \times
20^\circ$. 
The set of the HD\,101412 measurements covers the time interval of 2006--2009.
More details can be found at http://grb.fuw.edu.pl/. CCD photometry
was done without any filter, so that the results are similar to
a broad-band $R$ colour.
The star was  also observed over eight
nights in 2010 April employing the 0.5\,m reflector with the classical photometer in
SAAO in $\mathit{UBV}$, using a
fairly conventional single channel photometer with a Hamamatsu
R943--02 GaAs tube. We obtained 78 triads of measurements with an inner
accuracy of 9.4, 6.4, and 6.6\,mmag, respectively.

The very good initial estimate of the period allows us to describe the observed
periodic variations by a series of phenomenological models described
by a minimum number of free parameters, including the period $P$ and
the origin of epoch counting $M_0$.
The behaviour of the light curves in $V$ and $I$ is nearly the same. For
simplicity we assumed a similar behaviour also for light curves in $U$ and $B$
bands, while the light curve $R$ behaves differently. The periodic component in
light variations can then be described by means of periodic
functions $F(\vartheta)$ and $F_R(\vartheta_R)$:

\begin{eqnarray}\label{fce}
F(\vartheta,\beta_1,\beta_2)=\sqrt{1\!-\!\beta_1^2\!-\!\beta_2^2}\
\cos(2\,\pi\,\vartheta)+\,
\beta_1\cos(4\,\pi\,\vartheta)+\nonumber \\
\beta_2\hzav{\textstyle{\frac{2}{\sqrt{5}}}\,\sin(2\,\pi\,\vartheta)
-\textstyle{\frac{1}{\sqrt{5}}}\sin(4\,\pi\, \vartheta)},\\
F_R(\vartheta_R,\beta_3,\beta_4)=\sqrt{1\!-\!\beta_3^2\!-\!\beta_4^2}\
\cos(2\,\pi\,\vartheta_R)+\,
\beta_3\cos(4\,\pi\,\vartheta_R)+\nonumber \\
\beta_4\hzav{\textstyle{\frac{2}{\sqrt{5}}}\,\sin(2\,\pi\,\vartheta_R)
-\textstyle{\frac{1}{\sqrt{5}}}\sin(4\,\pi\, \vartheta_R)}.
\end{eqnarray}

The functions $F(\vartheta)$ and $F_R(\vartheta_R)$ are the simplest
normalised periodic function that represent the observed
photometric variations in detail. The phase of the brightness
extreme is defined to be 0.0, and the effective amplitude is
defined to be 1.0. The functions, being the sum of three terms, are
described by two dimensionless parameters $\beta_1,\,\beta_2$ and
$\beta_3,\,\beta_4$. $\vartheta$ and $\vartheta_R$ are the phase
function. Assuming linear ephemeris, we respectively obtain:

\begin{equation}
\vartheta=\zav{t-M_0}/P,\quad \vartheta_R=\vartheta - \Delta f_R,
\label{eq:ephem}
\end{equation}

\noindent
where $\Delta f_R$ is the phase shift of the basic minimum of the
function $F_R$ versus zero phase.

Periodic changes of magnitudes in $U,\,B,\,V,\,I$: $m_j(t)$ and
changes in $R$: $m_R(t)$ are given by the relations:

\begin{equation}
m_j(t)=\overline{m_j}+A\ F(\vartheta),\quad m_R(t)=\overline{m_R}+A\
F(\vartheta)+A_R\,F_R(\vartheta_R),\label{eq:mags}
\end{equation}

\noindent
where $A$ is the semiamplitude of light changes common to all bands and
$A_R$ is the amplitude of an additional component of light variability
being non-zero only in $R$. $\overline{m_j}$ and $\overline{m_R}$ are
mean magnitudes in the individual bands.

The periodic variations of the mean value of the longitudinal component of
the magnetic field intensity $\left<B_z\right>$ derived from all
lines can be well approximated by the simple cosinusoid:

\begin{equation}
\left<B_z\right>=\overline{\left<B_z\right>}+A_m\
\cos\hzav{2\,\pi\zav{\vartheta-\Delta f_m}},\label{eq:magn}
\end{equation}

\noindent
where $\overline{\left<B_z\right>}$ is the mean magnetic field
intensity, $A_m$ is the semiamplitude of the variations, and $\Delta f_m$
is the phase of the magnetic field minimum.

All 18 model parameters were computed simultaneously by a weighted
non-linear least-square method regression applied to the complete observational
material representing in total 3134 individual measurements. We
found the refined value of the rotational period
$P=42\fd076(17)$ and the origin of phase counting put at the $UBVI$
light maximum $M_0=2452797.9(8)$.

\begin{equation}
\mathrm{JD_{max}}=2\,452\,797.9(8) + 42\fd076(17) E.
\end{equation}

Parameters describing the functions $F$ and $F_R$ are:
$\beta_1=0.29(13)$,
$\beta_2=-0.49(10)$,
$\beta_3=0.36(16)$,
and $\beta_4=0.02(18)$,
the function $F_R$ primary minimum phase $\Delta f_{R1}=0.24(4)$,
the phase of the secondary minimum $\Delta f_{R2}=0.72(12)$,
the semiamplitude of light changes $A=-4.7(5)$\,mmag,
the semiamplitude $A_R=14.1(2.2)$\,mmag,
$\overline{V}=9.2660(14)$\,mag,
$\overline{B}=9.4501(13)$\,mag,
and $\overline{U}=9.6106(18)$\,mag.
The phase of the minimum of the mean projected intensity of the magnetic
field $\left<B_z\right>$, $\Delta f_m=-0.091(33)$, the mean value of
it: $\overline{\left<B_z\right>}=9\pm18$\,G, and the semiamplitude
$A_m=465\pm27$\,G.

\section{Appendix B: Non-periodic variations}

\begin{figure}
\centering
\includegraphics[width=0.45\textwidth]{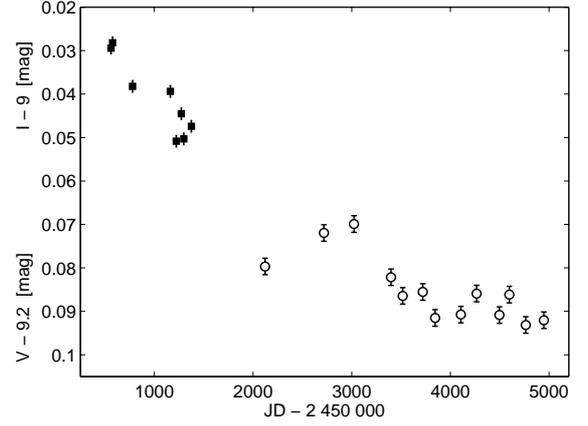}
\caption{Long-term changes in ASAS $V$ (open circles) and $I$ (full
squares) corrected for variations.}\label{fig:longterm}
\end{figure}

Relatively precise ASAS $V$ and $I$ data show an apparent
evolution of the mean values to the extent of several hundredths
of a magnitude over time scales of several years (see Fig.~\ref{fig:longterm}).
For this reason, we use detrended magnitudes, i.e.\ magnitudes where we
have removed the long term trends, in our search of periodic stellar variability.

An inspection of the light variation in $R$ plotted over the
rotational phase in 2006--2009 reveals dramatic seasonal
variations in the shape of the light curve. Unfortunately, the low
accuracy of the $R$ photometric measurements together with the observed
long-term instability and the large stochastic changes do not allow
us to analyse this phenomenon more deeply.
Using principal component analysis tools, we tested the possibility of seasonal changes in the
shape of the $V$ light curve in the first decade of this century.
We conclude that the $V$ light curve as defined by ASAS and SAAO
measurements was quite stable during this period.

Herbig Ae/Be stars are known to vary in a complex way.
The mechanism causing this variability is not well understood.
These more or less
stochastic variations amount up to several hundreds of a magnitude on
time scales from several minutes to tens of years (see
e.g.\ Ruczinski et al.\ \cite{slavek10}). The light changes of HD\,101412 in $V$ and $I$
presented in Fig.~\ref{fig:longterm} can be treated as
stochastic variations on the scale of tens of years. Nevertheless,
there is also evidence of such changes on much shorter time-scales.

\begin{figure}
\centering
\includegraphics[width=0.45\textwidth]{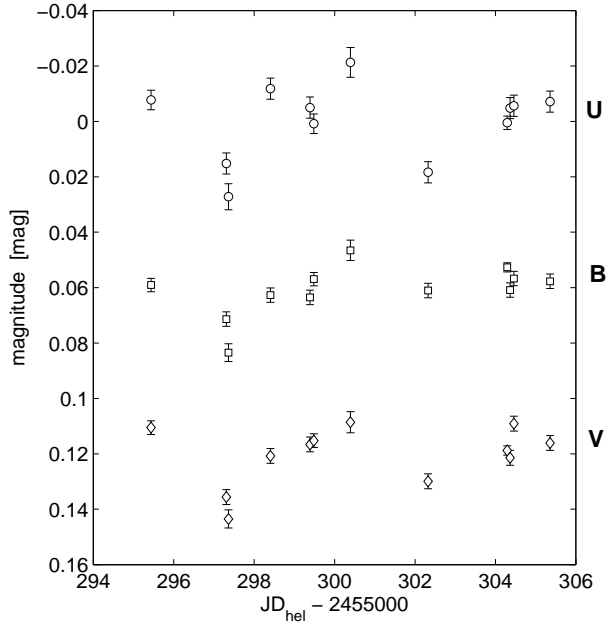}
\caption{Stochastic changes in our SAAO $UBV$ observations. Note
that the scatter in individual bands is highly
correlated.}\label{fig:UBV}
\end{figure}

Our own 78 $UBV$ photoelectric measurements in 2010 April in
SAAO were obtained in a total of 12 sets over eight nights. The duration of one
measurement set was not longer than 15 minutes and we assumed that
the brightness of the star did not change during the exposure. Then we can relatively
well estimate the typical inner uncertainty of one individual
measurement as 9.4, 6.4, and 6.6\,mmag in $U,\,B$, and $V$,
respectively. However, the means in particular sets in $U$, $B$, and
$V$, consisting of six individual measurements in the average, show an
additional scatter of 14, 10, and 11\,mmag (see Fig.~\ref{fig:UBV}).
We inferred that this scatter is due to stochastic changes on
the scale of hours or tens of hours. It should be noted that
the observed stochastic variations in particular colours are highly
correlated, whereas their amplitude is the largest in the $U$-band. Such a behaviour
is certainly somehow related to the physics of the mechanisms causing such
changes.

Because of the circumstance that we do not have at our disposal measurements in
$V$ and $I$ during one night, we estimate the inner accuracy of
these measurements implicitly. The mean scatter of detrended
ASAS data in $V$ and $I$ are 16 and 10\,mmag. Assuming that the
estimate of stochastic noise in $V$ of 11\,mmag is valid or smaller
for all ASAS $V$ measurements, we conclude, that the inner
uncertainty of one ASAS $V$ observation is smaller than 12\,mmag.
The ASAS estimates of uncertainty of $I$ observations should 
then be smaller by a factor of 0.6 compared to $V$ observations. The inner accuracy
should then be about 7\,mmag, as well as the measure of the stochastic
scatter. In all cases the stochastic scatter in $I$ is not allowed to
be larger than 10\,mmag. It means that the tendency of the decrease of
the stochastic scatter with the increase of the wavelength is supported
also by ASAS data.
A different amount of stochastic variation is expected in $R$ band, where
an additional variability mechanism is probably active. The mean scatter in
the $R$-band is about 56\,mmag. The inner uncertainty of Pi measurements
can be derived from magnitudes obtained in individual nights -- it
is estimated to be 49\,mmag. The stochastic scatter then can be
evaluated as 30\,mmag.

\end{document}